\documentclass[twocolumn,aps,prl,superscriptaddress,showpacs]{revtex4-1}
\usepackage{graphicx}
\usepackage{dcolumn}
\usepackage{bm}
\usepackage{color}
\usepackage{ulem}
\usepackage[english]{babel}
\begin{document}

\title{Nonlinear random optical waves: integrable turbulence, rogue waves and intermittency} 
\author{St\'ephane Randoux} 
\affiliation{Laboratoire de
  Physique des Lasers, Atomes et Molecules, UMR-CNRS 8523,
  Universit\'e de Lille, France} 
\author{Pierre Walczak}
\affiliation{Laboratoire de Physique des Lasers, Atomes et Molecules,
  UMR-CNRS 8523,  Universit\'e de Lille, France} 
\author{Miguel Onorato} 
\affiliation{Dipartimento di Fisica, Universit\`a degli Studi di Torino, 10125 Torino, Italy}
\affiliation{Istituto Nazionale di Fisica Nucleare, INFN, Sezione di Torino, 10125 Torino, Italy} 
\author{Pierre Suret}
\affiliation{Laboratoire de Physique des Lasers, Atomes et Molecules,
  UMR-CNRS 8523,  Universit\'e de Lille, France}

\begin{abstract}
We examine the general question of statistical changes experienced by ensembles of
nonlinear random waves propagating in systems ruled by integrable equations. 
In our study that enters within the framework of integrable turbulence, we specifically 
focus on optical fiber systems accurately described by the integrable one-dimensional 
nonlinear Schr\"odinger equation. 
We consider random complex fields having a gaussian statistics and an infinite extension 
at initial stage. We use numerical simulations with periodic boundary conditions and 
optical fiber experiments to investigate spectral and statistical
changes experienced by nonlinear waves in focusing and in defocusing propagation regimes.
As a result of nonlinear propagation, the power spectrum of the random wave broadens 
and takes exponential wings both in focusing and in defocusing regimes. Heavy-tailed 
deviations from gaussian statistics are observed in focusing regime while low-tailed 
deviations from gaussian statistics are observed in defocusing regime. After some transient 
evolution, the wave system is found to exhibit a statistically stationary state 
in which neither the probability density function of the wave field nor the spectrum 
change with the evolution variable. Separating fluctuations of small scale from fluctuations 
of large scale both in focusing and defocusing regime, we reveal the 
phenomenon of intermittency; i.e., small scales are 
characterized by large heavy-tailed deviations from Gaussian statistics, while the large 
ones are almost Gaussian. 
\end{abstract}



\maketitle

\section{Introduction}\label{Sec:intro}

The field of modern nonlinear physics has started with the pioneering work of 
Fermi and collaborators \cite{fermi1955studies} who studied a chain 
of coupled anharmonic oscillators, now known as the FPU system, with the aim of  understanding  the effect of nonlinearities in the process of thermalization. Their unexpected results, i.e. the observation of a recurrent behavior instead 
of the phenomenon of thermalization, triggered the work by Zabusky
and Kruskal \cite{zabusky1965} who  performed numerical simulations of the Korteweg de Vries 
equation (KdV), i.e. the long wave approximation of the 
FPU system, and made the fundamental discovery of solitons; such discovery lead to the development of 
a new field in mathematical physics that deals with integrable systems with an infinite number of 
degrees of freedom. Some years after the discovery, Zakahrov and Shabat \cite{zakharov72} found that 
the Nonlinear Schr\"odinger equation (1D-NLSE) is an integrable partial differential equations and has  
multi-solution just like the KdV equation. The following years were characterized 
by  the search of new integrable equations and the study of their mathematical properties and solutions.

 In the late seventies and early eighties, besides solitons on a zero background, a new class of solutions of the 1D-NLSE were found \cite{kuznetsov1977solitons,ma1979perturbed,peregrine1983water,akhmediev1985generation} that describe the instability of a finite coherent background. Those solutions are sometimes named ``breathers": 
the classical Akhmediev  \cite{akhmediev1985generation} and
the Peregrine  \cite{peregrine1983water} solutions breath just once in their life and 
they describe {\it in toto} the modulational instability process (also in its nonlinear stages). The Kuznetsov-Ma solution \cite{kuznetsov1977solitons,ma1979perturbed} is 
periodic in the evolution variable and the perturbation of the coherent state is never small. 

The phenomenon of rogue waves (RW)  in the ocean has  been known to
humans much before the discovery of integrability of partial
differential equations; however, only in the last fifteen years a
connection between those two fields has been made and it has been conjectured that the ``breather'' solutions
of the focusing 1D-NLSE  could be considered as rogue wave prototypes \cite{dysthe99, osborne00}.
This idea has been rapidly picked up in different scientific
communities \cite{kharif2009rogue,Akhmediev:09, Akhmediev:09b, Walczak:15,Dudley:14,Onorato:13,osborne2010nonlinear} and a new field with fresh ideas and old equations has started. A first important step was the reproduction of 
the breather solutions of the 1D-NLSE  in water wave tanks 
\cite{Chabchoub:11,Chabchoub:13} and in optical fibers \cite{Kibler:10,Kibler:12, Frisquet:13}.
In order to generate these coherent
structures in controlled lab experiments,  very specific and carefully-designed
{\it coherent} initial conditions have been considered. However, in nature such conditions are almost never encountered; wind generates ocean waves {\it via} a non trivial mechanism \cite{miles57,janssen91}  and 
the resulting wave field appears as a superposition of random waves characterized by Fourier spectra 
with a small, but finite, spectral bandwidth. This suggests that the problem of rogue waves should  
be investigated from a statistical point of view \cite{Onorato:04,  Erkintalo:10, Onorato:13,Walczak:15}. Indeed, one of the major question to be answered in the field of rogue waves concerns the determination of the probability density function (PDF) of the wave field for some given initial and boundary conditions. This is definitely not an easy task and, nowadays, given a nonlinear partial differential equation, there is no  systematic theory that allows one to determine the PDF of the wave field. 

The field of rogue waves has ``belong'' to oceanographers until the
the pioneering experiments with optical fibers described in \cite{Solli:07}. Since then, optical rogue waves 
have been  studied in various contexts such as supercontinuum
generation in fibers \cite{Solli:07, Solli:08,   Erkintalo:09,   Kibler:09, Mussot:09, Dudley:14}, propagation in optical fiber
described by the ``pure'' 1D-NLSE \cite{Walczak:15} or with higher
order dispersion \cite{Taki:10, Conforti:15}, laser filamentation
\cite{Kasparian:09}, passive cavities \cite{Montina:09, Conforti:15},
lasers \cite{Pisarchik:11,Bonatto:11,Lecaplain:12,Randoux:12} and
Raman fiber amplifiers \cite{Hammani:08}.

From the general point of view and beyond the question of rogue waves,
the field of nonlinear optics has then grown as a favorable laboratory
to investigate both statistical properties of nonlinear random waves and  hydrodynamic-like phenomena
\cite{Babin:07,Turitsyn:10, Churkin:10b, Michel:11,Turitsyna:13,Fatome:14,Picozzi:14}. Indeed, the field of incoherent dispersive waves resemble very much the classical field of fluid turbulence where, instead of waves, eddies interact with 
each other, giving birth to new eddies of different size. This mechanism is at the origin 
of the celebrated Kolmogorov  cascade of the three-dimensional turbulence which is 
characterized by a constant flux of energy within the  so called inertial range. 
A source and a sink of energy are required  in order to  maintain the cascade. Many years after such
concept was developed, it was found 
that the cascade is intermittent, i.e. the statistical properties of the velocity field 
vary with the scales, becoming less Normal for smaller scales, see for example \cite{Frisch:95}
for references.  In the light of the paper \cite{Randoux:14}, this idea will be discussed
in the present paper in the context of the dynamics of  incoherent 
waves ruled by the integrable 1D-NLSE. 
Such equation provides a bridge between nonlinear optics and hydrodynamics, see  \cite{chabchoub2015nonlinear} for a one to one comparison.  In particular, the focusing 1D-NLSE describes at leading order the physics of deep-water wave trains and it plays a central role in the study of rogue waves \cite{Onorato:01, Dudley:10, Onorato:13,  Akhmediev:13, Dudley:14}. Moreover, the focusing 1D-NLSE is the simplest partial differential equation that describes the  modulational instability phenomenon that is believed to be a fundamental mechanism for the formation of RW
\cite{Onorato:04,Onorato:13}. 

As mentioned,  such waves emerge in the ocean from the interplay of 
  incoherent waves in turbulent systems. The theoretical framework combining  a statistical
approach of random waves together with the property of integrability of the 1D-NLSE  is known as  {\it
  integrable turbulence}.    This emerging  field of research
recently introduced by V. Zakharov relies on the analysis of complex
phenomena found in nonlinear random waves systems described by an
integrable equation \cite{Zakharov:09,Zakharov:13,Pelinovsky:13,
  Agafontsev:14c,  Randoux:14, Suret:11}. Strictly speaking, the word ``turbulence'' is 
  not fully appropriate in the sense that the dynamics in Fourier space is not characterized 
  by a constant flux of a conserved quantity because the system is Hamiltonian 
  (no forcing and dissipation are included). For these integrable systems,  given an initial condition, the spectrum generally
  relaxes to a statistically stationary state that in general is different from the standard thermal equilibrium
characterized by the equipartition of energy. 
 The prediction of the spectra of such final state and its statistical properties is the objective of the integrable turbulence field.
 In the weakly nonlinear regime  \cite{Janssen:03}, starting with {\it incoherent} initial conditions in the 1D-NLSE, deviation from gaussian statistics has been predicted. In hydrodynamical numerical simulations performed with envelope equations and experiments made in water tanks, non gaussian statistics of the wave height has also been found to emerge from random initial conditions \cite{onorato2000occurrence,Onorato:04,Onorato:05}.

While in the water wave context the NLSE is only a crude (but reasonable) approximation of the original equations of motion, the field of nonlinear fiber optics is a promising field for the investigation of integrable
turbulence because  optical  tabletop ``model experiments''
accurately described by the 1D-NLSE can be performed  
\cite{Kibler:10,Kibler:12, Frisquet:13, Randoux:14}.   Despite the numerous works devoted  to optical RW,  the  generation of extreme events from purely  stochastic initial conditions in focusing 1D-NLSE model experiments remains  a crucial and open question \cite{Akhmediev:09b,Akhmediev:13,Agafontsev:14c, Dudley:14}.

In this paper, we review and extend a number of results recently obtained by the authors of this paper 
from optical fiber experiments  \cite{Randoux:14,Walczak:15} in the anomalous and normal dispersion regime. The dynamics of 
the waves in the considered fiber is described with high accuracy by the focusing and defocusing 1D-NLSE.
In the focusing regime, the idea is to 
implement optical fiber experiments conceptually analogous to the water
tank experiment described in \cite{Onorato:04} where waves with a finite spectral 
bandwidth and random phases are generated at one end of 
the tank and the evolution of the statistical properties of the wave field is followed along the flume.  Using an original
setup to overcome  bandwidth limitations of usual detectors,  we 
evidence  strong distortions of the statistics of nonlinear random
light characterizing the occurrence of optical rogue waves in integrable turbulence. 

In the defocusing regime, modulational instability is not possible and the evolution of 
incoherent waves does not lead to the formation of rogue waves. The statistics 
of wave intensity, initially following the  central limit theorem, changes along the fiber
resulting in a  decrease of  the tails of the PDF. This implies that the 
probability of finding a rogue wave is lower than the one described by linear theory.
 Implementing an optical filtering technique, we also report on the
 statistics of intensity of light fluctuations on different scales and
 we observe that the PDF of the wave
 intensity show tails that strongly depend on the scales. This reveals
 the phenomenon of intermittency, previously mentioned, that is
 similar to the one reported in several other wave systems, though
 fundamentally far from being described by an integrable wave
 equation. We also report new and original experimental results for partially coherent waves having a broad spectrum. We demonstrate in particular the emergence of strongly non gaussian statistics with low tailed PDF  in the defocusing  regime.
 
 The paper is organized as follows: numerical simulations of focusing and defocusing 1D-NLSE 
 are first considered and described in Sec. \ref{Sec:global_stat}. 
 Optical fiber experiments designed to investigate changes in the statistics 
 of random light fields and showing results in agreement 
 with simulations are presented in Sec. \ref{Sec:manip}.  
 In Sec. \ref{Intermittency}, we show that the integrable wave system under consideration 
 exhibit a phenomenon of intermittency both in focusing and defocusing regime. 
 In Sec. \ref{Conclusion}, we summarize our work and we discuss 
 open questions about integrable turbulence.

\section{Spatio-temporal, spectral and statistical features arising from nonlinear propagation of random waves in systems ruled by the integrable 1D-NLSE }\label{Sec:global_stat}

\subsection{General framework and description of the random initial condition}\label{Sec:CI}

Our study enters within the general framework of the integrable 1D-NLSE: 
\begin{equation}\label{nlse}
  i \psi_t +  \psi_{xx} +  2 \, \sigma \, |\psi|^2 \, \psi=0
\end{equation}
where $\psi(x,t)$ is the complex wave envelope. The parameter $\sigma$ 
determines the focusing ($\sigma=+1$) or defocusing ($\sigma=-1$) nature 
of the propagation regime. In nonlinear fiber optics, 
it is relatively easy to explore each of the two propagation regimes 
just by changing either the fiber or the wavelength of light \cite{Agrawal}. 
Eq. (\ref{nlse}) conserves
the energy (or Hamiltonian) $H=H_L+H_{NL}$  that has a nonlinear contribution
$H_{NL}=-\sigma \int | \psi(x,t) |^4 dx$ and a linear (kinetic) 
contribution $H_L= \int k^2 |\widehat{\psi}(k,t)|^2 dk$, the Fourier transform 
being defined as 
$\widehat{\psi}(k,t)=1/\sqrt{2\pi} \int \psi_0(x,t) e^{-ikx} dx$.
Eq. (\ref{nlse}) also conserves the number of particules (or power) 
$N=\int |\psi(x,t)|^2 dx$ and the momentum 
$P=\int k |\psi(k,t)|^2 dk$.

In our paper, the initial conditions that are used are 
\emph{non-decaying} random complex fields. 
We examine a situation that is very different 
from the problem of the Fraunhoffer diffraction of 
nonlinear spatially incoherent waves already considered in ref. 
\cite{Bass:87,Bromberg:10,Derevyanko:08,Derevyanko:12}.
In these papers, the nonlinear propagation of 
a speckle pattern of \emph{limited and finite} spatial 
extension is studied in focusing and defocusing media.
On the other hand we consider here continuous random waves of 
infinite spatial extension. This corresponds for instance to 
an experimental situation in which a partially coherent 
and continuous (i.e. not pulsed) light source of high 
power is launched inside a single-mode optical 
fiber \cite{Randoux:14,Walczak:15}. In numerical simulations, 
the random waves are confined in a 
box of size $L$ and periodic boundary conditions ($\psi(x=0,t)=\psi(x=L,t)$) 
are used to describe their time evolution \cite{Nazarenko}. 

The random complex field $\psi(x,t=0)=\psi_0(x)$ used as initial condition 
in this paper is made from a discrete sum of Fourier components : 
\begin{equation}\label{ini_field}
   \psi(x,t=0)=\psi_0(x)=\sum\limits_{n} \widehat{\psi_{0n}} e^{i n k_0 x}.
\end{equation}
with $\widehat{\psi_{0n}}=1/L \int_0^L \psi_0(x) e^{-ink_0x} dx$ and $k_0=2 \pi/L$.
The Fourier modes 
$\widehat{\psi_{0n}}=|\widehat{\psi_{0n}}|e^{i \phi_{0n}}$ 
are complex variables. In the random phase and amplitude (RPA)
model, generation of a random initial complex field is achieved by taking 
$|\widehat{\psi}_{0n}|$ and $\phi_{0n}$ as randomly-distributed 
variables \cite{Nazarenko}. 
Here, we will mainly use the so-called random phase (RP) model 
in which only the phases $\phi_{0n}$ of the Fourier modes are considered as 
being random \cite{Nazarenko}. In this model, the phase of each Fourier 
mode is randomly and uniformly distributed between $-\pi$ and $\pi$. 
Moreover, the phases of separate Fourier modes are not correlated so that 
$<e^{i\phi_{0n}}e^{i\phi_{0m}}>= \delta_{nm}$. In the previous expression, 
the brackets represent an average operation made over an ensemble 
of many realizations of the random process. $\delta_{nm}$ is the 
Kronecker symbol defined by $\delta_{nm}=1$ if $n=m$ and $\delta_{nm}=0$ 
if $n\neq m$.

With the assumptions of the RP model above described, 
the statistics of the initial 
field is homogeneous, which means that all statistical moments of the 
initial complex field $\psi_0(x)$ do not depend on $x$ 
\cite{Picozzi:07,Picozzi:14}. 
The power spectrum $n_0(k_n)$ of the random field $\psi_0(x)$ then reads as : 
\begin{equation}\label{power_spectrum}
<\widehat{\psi_{0n}}\widehat{\psi_{0m}}>=n_{0n} \, \delta_{nm}=n_0(k_n). 
\end{equation}
with $k_n=n\,k_0$. In the limit where $L \rightarrow \infty$, 
the frequency separation between two neighboring frequency components 
$k_n$ and $k_{n+1}$ tends to zero and the discrete 
spectrum $n_0(k_n)$ becomes a continuous spectrum $n_0(k)$.

The RP model is often used in the contexts of hydrodynamics 
where the power spectrum $n_0(k)$ is given by the 
so-called JONSWAP spectrum \cite{Islas:05,Onorato:01} 
It has also been used in optics where simple gaussian or 
sech profiles are often used for the function $n_0(k)$
\cite{Suret:11,Randoux:14,Walczak:15}. 

\begin{figure}[htb]
\centerline{\includegraphics[width=8cm]{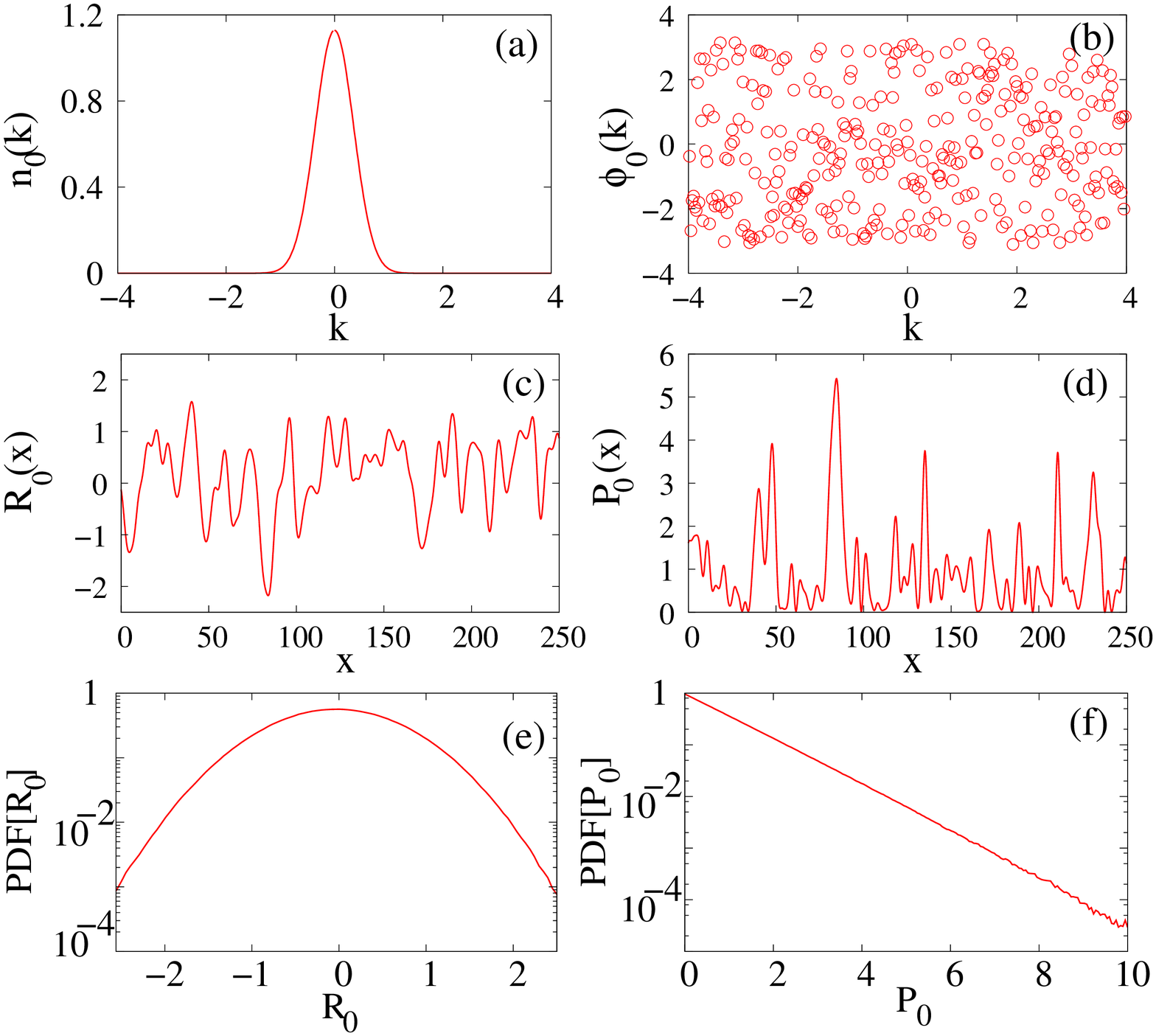}}
\caption{
(a) Gaussian power spectrum $n_0(k)$ of the initial condition 
such as defined by Eq. (\ref{gaussian_ci}) ( $\Delta k=0.5$, 
$n_0=1.129$). (b) Spectral 
distribution of the phases $\phi_0(k)$ of the Fourier modes used 
to compute the random initial field $\psi_0(x)$. (c) Real part 
$R_0(x)=\Re{(\psi_0(x))}$ and (d) power $P_0(x)=|\psi_0(x)|^2$ 
of the random field computed from spectra shown in (a), (b). 
(e) PDF of $R_0(x)$ and (f) PDF of $P_0(x)$ showing that 
the initial condition has a gaussian statistics. 
} 
\end{figure}

In this Section, we consider a random complex initial field 
having a gaussian optical power spectrum that reads 
\begin{equation}\label{gaussian_ci}
n_0(k)=n_0 \, \exp \left[- \left( \frac{k^2}{\Delta k^2} \right) \right]
\end{equation}
where $\Delta k$ is the half width at $1/e$ of the power spectrum. 
Fig. 1 shows a typical example of a partially coherent complex field generated
using the RP model and a power spectrum given by Eq. (\ref{gaussian_ci}). 
Fig. 1(b) shows that the values of the spectral phases $\phi_0(k)$ 
are randomly distributed between $-\pi$ and $\pi$. Fig. 1(c) shows 
the random evolution of 
the real part $R_0(x)=\Re{(\psi_0(x))}$ of the initial field that is computed 
from the spectra shown in Fig. 1(a) and 1(b). We do not present here 
the evolution of the imaginary part $I_0(x)=\Im{(\psi_0(x))}$ of 
the initial field because it is qualitatively very comparable 
to what is shown in Fig. 1(c). However it is important to notice that
the RP model produces a random field having real and imaginary parts 
that are statistically independent, i.e. 
$<R_0(x)I_0(x)>=0 \, \, \, \forall \, x$.
The spatial evolution of the power $P_0(x)=|\psi_0(x)|^2=R_0^2(x)+I_0^2(x)$ 
of the random initial field is shown in Fig. 1(d). 
Note that the numerical values of the parameters $n_0$ and 
$\Delta k$ have been chosen in such a way that the number 
of particules $N=1/L \int_0^L |\psi(x,t)|^2 dx$ is equal to unity. 

As previously described, the random complex field $\psi_0(x)$ used as 
initial condition is produced from the linear superposition 
of a large number of independent Fourier modes having randomly 
distributed phases. As stated by the central limit theorem, 
the statistics of the random process produced from such a superposition 
follows the normal law. The RP model thus produces a complex field having 
quadratures that are statistically independent and that have the 
same gaussian statistics. It is straightforward to prove that the 
statistics of power fluctuations $P_0$ follows the exponential 
distribution \cite{Mandel_wolf, Agafontsev:14c}. Note that the 
PDF for the fluctuations of the 
amplitude $A_0(x)=|\psi_0(x)|$ is given by the Rayleigh 
distribution defined by $PDF[A_0/<A_0>]=A_0/<A_0> \, . \, exp(-A_0/<A_0>)$ 
\cite{Mandel_wolf, Agafontsev:14c}

In order to illustrate these statistical features from numerical 
simulations, we have performed the analysis of the statistical 
properties of the complex field generated from the RP model
by producing an ensemble of $10^4$ realizations of the 
random initial field.
From this ensemble, it is in particular possible to compute 
the PDF for the fluctuations of 
$R_0(x)=\Re{(\psi_0(x))}$ and of $P_0(x)=|\psi_0(x)|^2$. Fig. 1(e) 
shows that the PDF of $R_0(x)/<R_0(x)>$ is gaussian. The PDF of 
the normalized imaginary part $I_0(x)/<I_0(x)>$ of the complex field 
is not shown here but it is rigorously identical to the PDF of 
$R_0(x)/<R_0(x)>$. As it is illustrated in Fig. 1(f), 
the PDF of the power is given by the exponential distribution, i.e. 
$PDF[P_0/<P_0>]=exp(-P_0/<P_0>)$.

\subsection{Focusing regime}\label{Sec:focusing}

In this Section, we consider the focusing regime ($\sigma=+1$) 
and we use numerical simulations of Eq. (\ref{nlse})
to investigate the propagation of a partially coherent wave
generated at $t=0$ by the random complex field described in 
Sec. \ref{Sec:CI}. We will consider the spatio-temporal 
dynamics of the partially coherent wave and we will also discuss 
spectral and statistical changes occurring in time. 

Our numerical simulations have been performed by using a pseudo-spectral
method working with a step-adaptative algorithm permitting to reach a specified 
level of numerical accuracy. The numerical simulations are performed 
by using a box of size $L=257.36$ that has been discretized by using $4096$ 
points. Statistical properties of the random wave are computed 
from an ensemble of $10^4$ realizations of the random initial condition. 

Fig. 2(a) shows the spatio-temporal evolution 
of the partially-coherent wave seeded by a random initial field 
having properties that are described in Sec.\ref{Sec:CI} and 
that are synthesized in Fig. 1. At the initial stage of the nonlinear 
evolution ($t \sim 0$), the power $|\psi(x,t)|^2$ of the wave 
is slowly and randomly modulated. The spatial scale of the fluctuations 
of $|\psi(x,t)|^2$ at $t \sim 0$ is determined by $\Delta k$ 
and it is typically around $10$ in the simulations shown in Fig. 2
(see also Fig.1(d)). Numerical simulations show that series of peaks emerge
from the random initial condition. The density of these peaks 
is higher is those regions of space where the complex field exhibits 
high peak power fluctuations at the initial time $t=0$, see e. g. the region 
where $x \in [-50,-40]$ in Fig. 2 and in Fig. 3(a),(c),(e). 
While the localized peaks shown in Fig. 2(a) drift with small 
velocities in the $(x,t)$ plane, 
their peak power increases with time. This increase of the peak power 
of the localized structures goes simultaneously with a reduction of 
their spatial width. The typical spatial scale of the power fluctuations 
evolves from a value of $\sim 10$ at initial stage ($t=0$) to the healing 
length of $\sim 1$ at long evolution time ($t>10$) (see Fig. 3(a) , (c), (e)). 

A clear signature of the change in the fluctuation scale can be observed in the Fourier space. 
As shown in Fig. 3(b), 3(d) and 3(f), the power 
spectrum $|\widehat{\psi}(k,t)|^2$ of the random wave
is indeed found to significantly broaden with time. 
Let us emphasize that the power spectrum of the wave 
broadens while always keeping exponentially decaying wings. 

\begin{figure}[htb]
\centerline{\includegraphics[width=8cm]{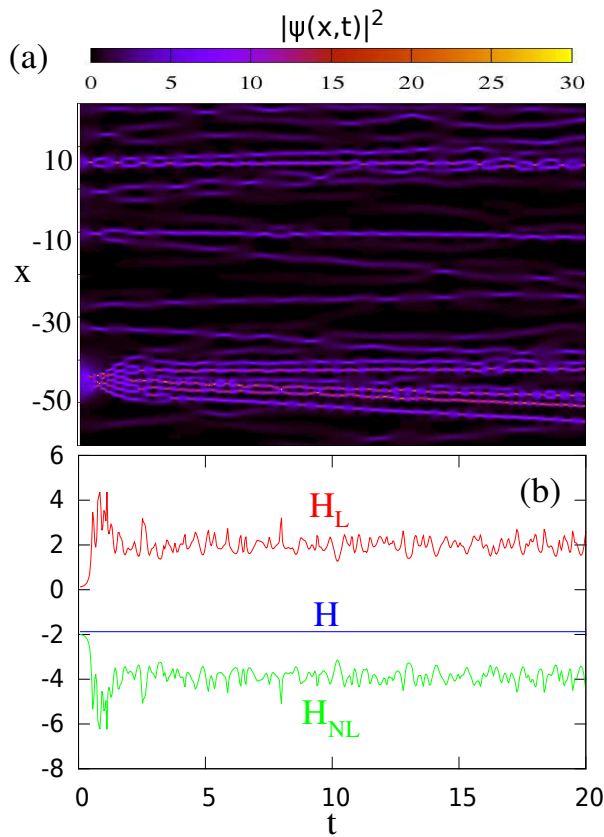}}
\caption{Numerical simulations of Eq. (\ref{nlse}) in focusing regime 
($\sigma=+1$). (a) Spatio-temporal evolution of the power 
$|\psi(x,t)|^2$ of the wave while starting from the random 
complex field having a gaussian power spectrum and a gaussian 
statistics, see Fig. 1. (b) Corresponding time evolution of linear (kinetic)
energy $H_L$ and of nonlinear energy $H_{NL}$.
} 
\end{figure}

The evolution of the typical space scales and peak powers of the random 
fluctuations that is described above 
goes together with a process of energy balance between linear and nonlinear effects. 
Fig. 2(b) shows the evolutions in time of the linear (kinetic)
energy $H_L$ and of the nonlinear energy $H_{NL}$ 
that are associated with the spatio-temporal evolution plotted 
in Fig. 2(a). The wave system is initially placed in a highly nonlinear
regime in which the nonlinear energy is one order of magnitude 
greater than the linear energy ($|H_{NL}| \simeq 10 |H_L|$). As 
a result of nonlinear propagation, linear and nonlinear 
effects come into balance and after a short transient evolution, 
the wave system reaches a state in which linear and 
nonlinear energies have the same order of magnitude 
($|H_{NL}|\simeq 2|H_L|$, see Fig. 2(b) ).

\begin{figure}[htb]
\centerline{\includegraphics[width=8cm]{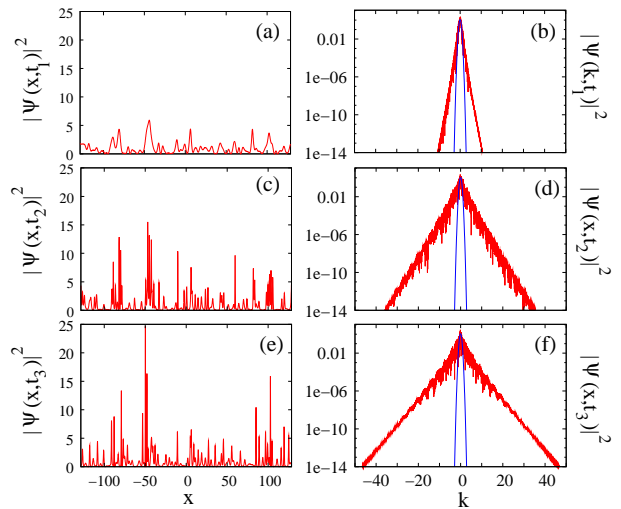}}
\caption{Numerical simulations of Eq. (\ref{nlse}) in focusing regime 
($\sigma=+1$). Spatial evolution of the power $|\psi(x,t)|^2$ 
of the random wave at times (a) $t_1=0.24$, (c) $t_1=2.44$ , (e) $t_1=16.04$. 
Power spectra (red lines) $|\psi(k,t)|^2$ of the random wave 
at times (b) $t_1=0.24$, (d) $t_1=2.44$ , (f) $t_1=16.04$. 
The spectra plotted in blue lines represent the gaussian power spectrum
of the random initial complex field defined by Eq. (\ref{power_spectrum})
and Eq. (\ref{gaussian_ci}).
} 
\end{figure}

It is interesting to compare 
the spatio-temporal evolution shown in Fig. 2(a) with the one 
shown in Fig. 2(a) of ref. \cite{Toenger:15}. In ref. \cite{Toenger:15}, 
the authors study the nonlinear evolution in space and time 
of a condensate perturbated by a small noise, i. e. $\psi_0(x)=1+\eta(x)$
where $|\eta(x)|<<1$ is a small noise with broad spectrum. 
As shown in Fig. 2(a) of ref. \cite{Toenger:15}, the fact that 
there is only a weak random modulation of the initial condition gives rise 
to a spatiotemporal diagram in which localized structures are 
distributed in space and time in a way that is more regular than 
the random pattern shown in Fig. 2(a) of this paper. Starting from our initial 
condition with a broad gaussian spectrum, there are some wide regions of 
space in which no localized structures are observable while there 
are some other regions of space including many localized structures. 
As shown in ref. \cite{Walczak:15}, the localized structures 
emerging from a complex field having initially a 
gaussian statistics can be locally fitted by some analytical 
functions corresponding to solitons on finite background, 
such as e. g. the Peregrine soliton. Fitting procedures 
implemented in the numerical work presented in 
ref. \cite{Toenger:15} have shown that many 
solitons on finite background can be also found while 
seeding the wave system from a condensate perturbated by a 
small noise. 

Despite localized structures looking like solitons on 
finite background can be observed while starting from those 
two different random initial conditions, significantly 
different statistical features are observed at long evolution time.
It has been shown in ref. \cite{Agafontsev:14c} that 
gaussian statistics emerges from the nonlinear 
evolution of the noisy condensate. 
As shown in Fig.4(a), heavy-tailed deviations from gaussianity 
are contrarily found to emerge from a random complex field having 
initially a gaussian statistics \cite{Walczak:15}.
It is an open question to understand how the interplay among localized 
structures does not produce the same statistics at long evolution 
time while starting from different noisy initial conditions.

Rational solutions of the 1D-NLSE such as Akhmediev breathers are
considered as prototype of rogue waves \cite{Akhmediev:09,Akhmediev:09b}. 
These coherent structures have been generated in optical fiber experiments
\cite{Kibler:10,Kibler:12} and in hydrodynamical experiments 
\cite{Chabchoub:11}. The interaction of solitons \cite{Pelinovsky} 
and the collision of breathers have been studied theoretically and 
experimentally  \cite{Frisquet:13}. Our work points out the relevance 
of these works and the need to extend it in the context of random 
nonlinear waves. In particular, it is an open question to understand 
the mechanisms of the emergence of coherent structures from the two 
different initial conditions, i.e. the plane wave with small noise on one hand 
and the random wave computed from the RP model on the other hand.

Fig. 4 shows the time evolution of the PDF of $|\psi(x,t)|^2$. 
The nonlinear random field has a statistical evolution in which 
the PDF of power fluctuations continuously moves from the exponential
distribution (plotted in red line in Fig. 4) to the heavy-tailed 
distribution plotted in magenta line in Fig. 4. 
For $t>10$, the wave system reaches a statistical stationary state 
in which the PDF no longer changes with time \cite{Walczak:15}.  
This statistical stationary state is determined by the interaction 
of coherent nonlinear structures such as for instance 
Akhmediev breathers, Kuznetsov-Ma solitons and also linear dispersive 
radiation \cite{Pelinovsky:13,Efimov:05,Bohm:06}. It is now an open 
question to determine the mechanisms in integrable turbulence that
lead to the establishment of a stationary state in which
statistical properties of the wave system do not change in time. 
Tools from the inverse scattering theory could be used to 
investigate this question of fundamental importance 
\cite{Bass:87,Gennadyprl:05,Bohm:06,Derevyanko:08,Derevyanko:12}

\begin{figure}[htb]
\centerline{\includegraphics[width=6cm]{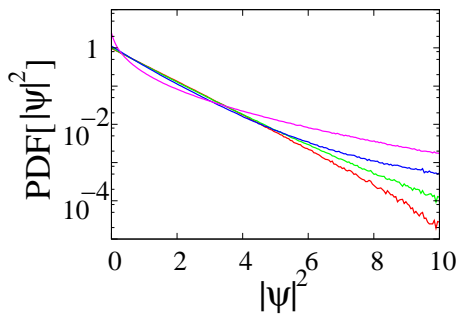}}
\caption{Numerical simulations of Eq. (\ref{nlse}) in focusing regime 
($\sigma=+1$). PDF of the power $|\psi(x,t)|^2$ at time $t=0$ (red line), 
$t=0.2$ (green line), $t=0.4$ (blue line), $t=10$ and $t^*=20$ (magenta line).
The PDF is stationary from $t\sim 10$, i.e. the PDF plotted in magenta line 
does not change with time for $t>10$.
} 
\end{figure}

\subsection{Defocusing regime}\label{Sec:defocusing}

In this Section, we consider the defocusing regime ($\sigma=-1$) 
and we use numerical simulations of Eq. (\ref{nlse})
to investigate the propagation of a partially coherent wave
generated at $t=0$ by a random complex field identical to  
the one used in Sec. \ref{Sec:focusing}.

\begin{figure}[htb]
\centerline{\includegraphics[width=8cm]{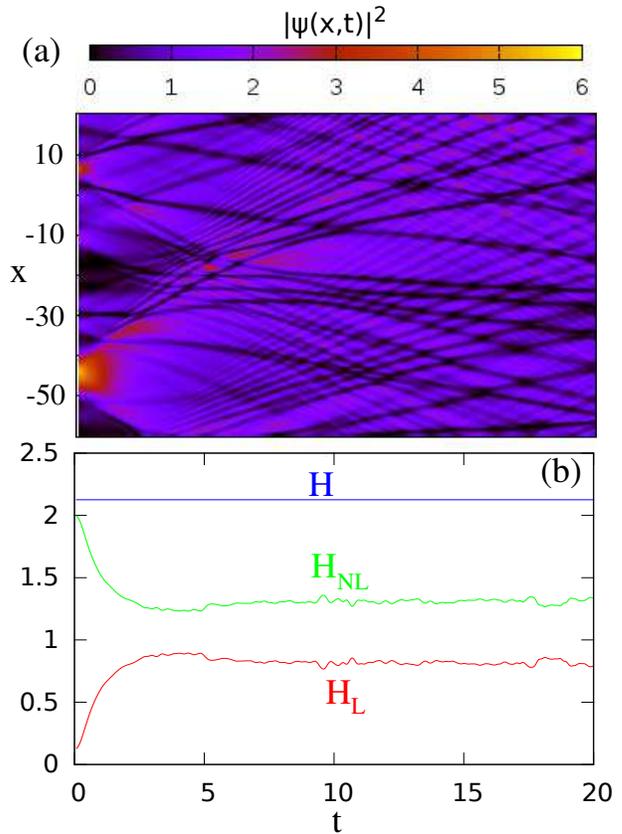}}
\caption{Numerical simulations of Eq. (\ref{nlse}) in defocusing regime 
($\sigma=-1$). (a) Spatio-temporal evolution of the power 
$|\psi(x,t)|^2$ of the wave while starting from the random 
complex field having a gaussian power spectrum and a gaussian 
statistics, see Fig. 1. (b) Corresponding time evolution of linear (kinetic)
energy $H_L$ and of nonlinear energy $H_{NL}$.
} 
\end{figure}

Fig. 5(a) shows the spatio-temporal evolution 
of the partially-coherent wave seeded by a random initial field 
identical to the one used in Fig. 2(a). Spatiotemporal 
features emerging from the nonlinear propagation in defocusing 
regime drastically contrast with those found in the focusing 
regime. Instead of bright localized structures, we now 
observe the emergence of dark localized structures propagating 
at various speeds in the $(x,t)$ plane. Fig. 6(a) and 6(b) show
that the initial stage of nonlinear evolution is now characterized 
by a fast decay of the peaks of highest intensities, see e. g. the region 
where $x \in [-50,-40]$. During the initial evolution of the 
random wave, the leading and trailing edges of peaks of highest 
intensities strongly sharpen. This leads to some gradient catastrophes
which are regularized by the generation of dispersive shock
waves (DSWs) \cite{Fatome:14,Conforti:13,Conforti:14,Moro:14,Gennady:05}. 
As in the focusing case, the typical spatial scale of the random fluctuations decreases 
from $\sim 10$ at the initial stage ($t=0$) to the healing length of 
$\sim 1$ at long evolution time ($t>10$). The stochastic evolution
shown in Fig. 6(c) is determined by the interaction 
of nonlinear coherent structures such as dark solitons or DSWs and 
of linear radiation. 

Fig. 6(b), 6(d) and 6(f) show that nonlinear propagation in defocusing
regime induces a spectral broadening of the random wave. This spectral 
broadening phenomenon is quantitatively less pronounced than the one 
observed in the focusing regime, see Fig. 3(b), 3(d), 3(f). However
the power spectrum of the wave broadens while always keeping 
exponentially decaying wings, as in the focusing regime.

\begin{figure}[htb]
\centerline{\includegraphics[width=8cm]{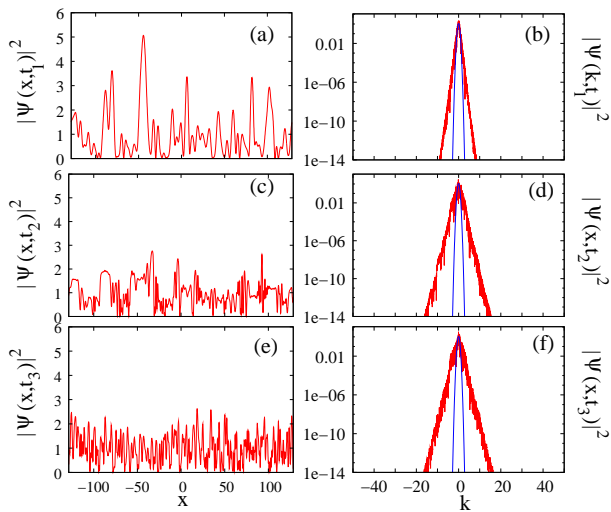}}
\caption{Numerical simulations of Eq. (\ref{nlse}) in defocusing regime 
($\sigma=-1$). Spatial evolution of the power $|\psi(x,t)|^2$ 
of the random wave at times (a) $t_1=0.24$, (c) $t_1=2.44$ , (e) $t_1=16.04$. 
Power spectra (red lines) $|\psi(k,t)|^2$ of the random wave 
at times (b) $t_1=0.24$, (d) $t_1=2.44$ , (f) $t_1=16.04$. 
The spectra plotted in blue lines represent the gaussian power spectrum
of the random initial complex field defined by Eq. (\ref{power_spectrum})
and Eq. (\ref{gaussian_ci}).
} 
\end{figure}

The spatiotemporal evolution shown in Fig. 5(a) and in Fig. 6(a), (c), (e)
goes together with a process of energy balance between linear and nonlinear effects. 
Fig. 5(b) shows the time evolutions of the linear (kinetic)
energy $H_L$ and of the nonlinear energy $H_{NL}$ 
that are associated with the spatio-temporal evolution plotted 
in Fig. 5(a). The wave system is initially placed in a highly nonlinear
regime in which the nonlinear energy is one order of magnitude 
greater than the linear energy ($H_{NL}\sim 10 H_L$). As in the focusing regime,
linear and nonlinear effects come into balance and after a short transient evolution, 
the wave system reaches a state in which $H_{NL}$ and $H_L$ have the same order of
 magnitude, see Fig. 5(b). 

\begin{figure}[htb]
\centerline{\includegraphics[width=6cm]{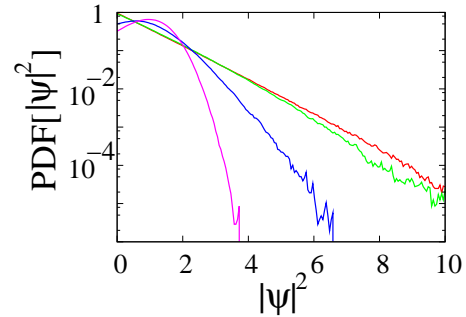}}
\caption{Numerical simulations of Eq. (\ref{nlse}) in defocusing regime 
($\sigma=-1$). PDF of the power $|\psi(x,t)|^2$ at time $t=0$ (red line), 
$t=0.2$ (green line), $t=1$ (blue line), $t=10$ and $t^*=20$ (magenta line).
The PDF is stationary from $t\sim 10$, i.e. the PDF plotted in magenta line 
does not change with time for $t>10$.
} 
\end{figure}

In the defocusing regime, the nonlinear random 
field has a statistical evolution in which 
the PDF of power fluctuations continuously moves from the exponential
distribution (plotted in red line in Fig. 7) to the low-tailed 
distribution plotted in magenta line in Fig. 7. 
As in the focusing regime, the wave system exhibits a 
statistical stationary state and the PDF computed at $t=10$ 
(magenta line in Fig. 7) does not change anymore with time \cite{Randoux:14}. 
This statistical stationary state is determined by the interaction 
of coherent nonlinear structures such as for instance 
dark solitons, dispersive shock waves and also linear
radiation. As for the focusing regime, it is an open 
question to determine the mechanisms in integrable turbulence that
lead to the establishment of a stationary state in which
statistical properties of the wave system do not change in time. 
Tools of the inverse scattering transform could be of interest
for the investigation of this question \cite{Fratolocchi:08}.

\section{Optical fiber experiments in focusing and in defocusing 
propagation regimes}\label{Sec:manip}

Optical fiber experiments provide versatile and powerful tabletop
laboratory to investigate the complex dynamics of 1DNLSE,
hydrodynamic-like phenomena and the statistical properties of nonlinear
random waves \cite{Babin:07,Turitsyn:10, Gorbunov:14,Turitsyna:13,Chabchoub:13,Fatome:14,Picozzi:14,Solli:07, Dudley:14}.

One of the most critical constraint of these experiments is the finite
spectral bandwidth of usual detectors. The typical response time of
the fastest detector and oscilloscope is several tens of picoseconds.
On the other hand, with the usual parameters of standard experiments
using optical fibers, the typical ``healing time'' scale
characterizing the equilibrium between the nonlinearity and the
dispersion is around one picosecond \cite{Walczak:15}. For this
reason, the picosecond is also the order of magnitude of the time
scale associated to the power fluctuations of partially coherent fiber lasers
\cite{Babin:07,Turitsyn:10,Walczak:15b}.

As a consequence  spectral filters are
therefore often used to reveal extreme events in time-domain
experiments \cite{Solli:07,Erkintalo:09, Randoux:12, Randoux:14}. In
the case of pulsed experiments, it is possible to evidence
shot-to-shot spectrum fluctuations with a dispersive Fourier transform
measurement \cite{Solli:07,Jalali:10, Wetzel:12,Goda:13}. To the best
of our knowledge, up to our recent works
\cite{Walczak:15,Walczak:15b}, the {\it accurate and well-calibrated }
measurement of  the PDF characterizing
temporal fluctuations of the power of {\it random light} with time
scale of the order of picosecond had never been performed.\\

 \begin{figure}[h]
\includegraphics[width=7cm]{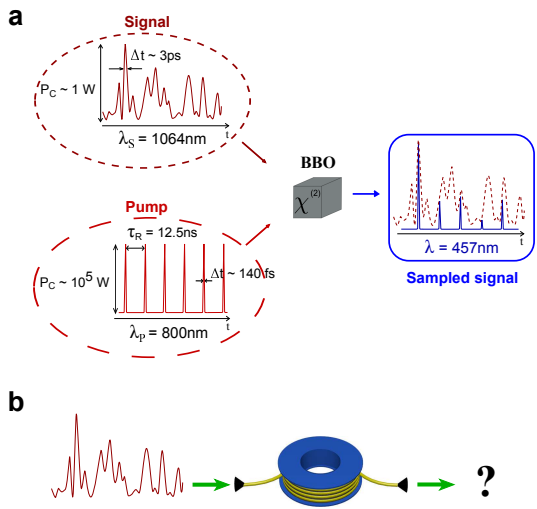}
\caption{ {\bf Experimental measurement of the statistics of
    random light} {\bf a. Principle}. Optical sampling of the
  partially-coherent wave  fluctuating with time (the signal) is
  achieved from sum frequency generation in a second order
  ($\chi^{(2)}$) crystal.  Blue pulses are generated at $\lambda=457$nm from
  the interaction of the signal with periodic femtosecond pump pulses
  inside a $\chi^{(2)}$ crystal. PDF is computed from the peak powers
  of the blue pulses.
 {\bf b. Nonlinear propagation in optical fiber} The initial partially
coherent light is launched inside a single mode optical fiber (either
in the focusing (fiber 1) or defocusing (fiber 2) regime of dispersion. The
statistics and the spectrum is measured before and after the
propagation in the fiber.}
\label{fig:setupOS}
\end{figure}

We have developed an original setup  based
on asynchronous optical sampling (OS) which allows the precise measurement of
statistics of random light rapidly fluctuating with time (see Fig. \ref{fig:setupOS}).

The ``random light''  under investigation is a partially coherent wave
and it is called the ``signal''. The signal is optically sampled with
 140fs-pulses and the PDF is computed from the samples.   The optical
 sampling is obtained from   the second order nonlinearity $\chi^{(2)}$ in a BBO
 crystal. The sum-frequency generation (SFG) between  the signal at
 $\lambda_S=1064$nm and short ``pump'' pulses having  a central
 wavelength  $\lambda_P=800$nm provide samples of the signal at a wavelength $\lambda=457$nm.

In our experiments, the linearly polarized partially coherent wave
is emitted by a ``continuous'' wave (cw) Ytterbium fiber laser  at
$\lambda_S=1064$nm. This cw laser
emits numerous (typically $10^4$) uncorrelated longitudinal modes. The
reader can refer to \cite{Walczak:15} for the details of the
experimental setup and of the statistics measurement procedure.

In this paper we present results obtained with two different fibers (fiber 1
and 2) having opposite sign of the group velocity dispersion at the
wavelength of the signal  $\lambda_S=1064$nm. The fiber 1 is a 15m-long
highly nonlinear photonic crystal fiber (provided by Draka France
company) with a nonlinear third order coefficient
$\gamma\simeq50$W$^{-1}$km$^{-1}$ and a group velocity dispersion
coefficient  $\beta_2\simeq-20ps^2/km$. The fiber 2 is a 100m-long
polarization maintaining fiber with a nonlinear third order coefficient
$\gamma\simeq 6$W$^{-1}$km$^{-1}$ and a group velocity dispersion
coefficient  $\beta_2\simeq+20ps^2/km$. We launch a mean power
$<P>=0.6$W in the experiments performed with the fiber 1 (focusing
case) and a mean power $<P>=4.$W in the experiments performed with the
fiber 2 (defocusing case). Note that the results obtained with the fiber 1 have
been presented in detail in \cite{Walczak:15} whereas the results
obtained with the fiber 2 are new.\\

\begin{figure}[h]
\includegraphics[width=8.cm]{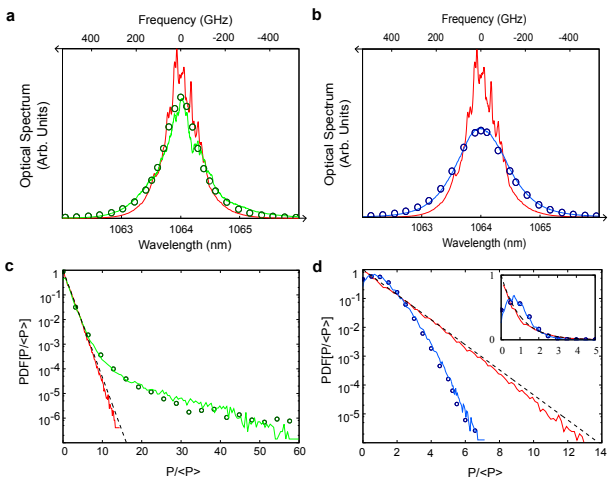}
\caption{ {\bf Experiments.} {\bf a,b :} Optical spectra. Input spectrum
  (red line). Spectrum at the output of fiber 1 (green line) and fiber
  2 (blue line). The circles represents optical spectra computed from
  numerical simulation of 1D-NLSE. {\bf c, d : }  PDF 
  of normalized optical power  $P/\langle P\rangle$
  plotted in logarithmic scale.  
  Normalized exponential distribution  $PDF[P/<P>]=\exp(-P/<P>)$
  (black dashed line). PDF of the input random light (red line). PDF
  of the light at the output of the fiber 1 (green solid line) and at
  the output of the fiber 2 (blue solid line). The dashed lines
  represent the PDF computed from numerical simulation of 1D-NLSE
  with the parameters of the experiments. The inset represents the
  same PDF in linear scale. }
\label{fig:expPDF}
\end{figure}

We first measure the PDF at the output of the laser. In all
experiments presented in this letter, the mean output power of the
Ytterbium laser is fixed at $\langle P\rangle=10$W. At this operating point, the
statistics of the partially coherent wave follows the normal law.
Indeed, as plotted in red in Figs. \ref{fig:expPDF}.c and
\ref{fig:expPDF}.d, the PDF of the normalized power $P/\langle
P\rangle$ is very close to the exponential function. Assuming that the real part and the imaginary
parts are  statistically independent, this exponential distribution of
power corresponds to a gaussian statistics of the field. It
is important to note that the  dashed black lines in Figs.
\ref{fig:expPDF}.c and \ref{fig:expPDF}.d are not a fitted
exponential function but  represent the exact 
normalized  $PDF[P/\langle P\rangle]=\exp(-P/\langle P\rangle)$.
To the best of our knowledge, PDF of so rapidly fluctuating optical
signals has never been {\it quantitatively} compared to the normalized exponential distribution. 

We use the output of the laser as a random source and we launch
the partially coherent signal into optical fibers 1 and 2.
Experiments have been carefully designed to be very well described by
the 1D-NLSE. In particular, the signal wavelength $\lambda_s=1064$nm is
far from the zero-dispersion wavelength ($\lambda_0$$\simeq 970$nm for
the fiber 1 and $\lambda_0 > 1300$nm for the fiber 2). Moreover the optical spectral widths (see Figs. \ref{fig:expPDF}.a and
\ref{fig:expPDF}.b)  remain sufficiently narrow to neglect stimulated
Raman scattering (SRS)   and high-order dispersion effects.  The
linear losses  experienced by optical fields  in single pass in the
fibers are neglictible.  These total losses are around $0.3\%$  in the fiber
1 and around around $2.5\%$ in the fiber 2.\\

 Despite the broadening of the optical spectrum is of nearly the same
 importance in focusing and in defocusing regimes (see Figs.
 \ref{fig:expPDF}.a and \ref{fig:expPDF}.b)  our experiments reveal
 that the distortions of the statistics  of the random waves strongly
 depend  on the sign of  the group  velocity dispersion coefficient.
 
The experiments performed in the {\it focusing} regime (fiber 1)
reveal the occurrence of numerous extreme events (RW) (see green curve
in Fig. \ref{fig:expPDF}.c).  The
comparison between the initial PDF (see red line in Fig.
\ref{fig:expPDF}.c) and the
output PDF (see green curve \ref{fig:expPDF}.c) shows an impressive change in
the statistical distribution of optical power. The initial field
follows the normal law and its PDF is an exponential function whereas
the output PDF of optical power exhibits a strong heavy-tail.

On the other hand, the PDF experimentally measured at the output of
fiber 2 in the defocusing regime exhibits a very low tail (see Fig.
\ref{fig:expPDF}.d). Light fluctuations of a high power are found with
a probability that has been strongly reduced as compared to the normal
law. Moreover, contrary to the initial exponential distribution,
the most probable value for the power is  not the zero value (see inset of
Fig. \ref{fig:expPDF}.d) ).

Numerical simulations show that experiments presented above  are very
well described by the integrable 1D-NLSE. The initial conditions are
computed from the random phase assumption as in the section
\ref{Sec:CI}. We have performed Monte Carlo simulations with ensemble
average over thousands of realizations. We integrate the 1D-NLSE with
experimental parameters :

\begin{equation}
  \label{eq:NLS1D}
  i\frac{\partial \psi}{\partial z}=\frac{\beta_2}{2}\frac{\partial^2
    \psi}{\partial t^2}-\gamma|\psi|^2\psi
\end{equation}

where $\beta_2$ is the  group velocity dispersion coefficient and
$\gamma$ is the effective Kerr coefficient. Optical spectra and PDFs
computed from the numerical integration of the 1D-NLSE   are in
quantitative agreement with experiments both in the focusing and in
the defocusing cases  (see dashed green and blue lines in Fig.
\ref{fig:expPDF}).

\begin{figure}[h]
\includegraphics[width=8.cm]{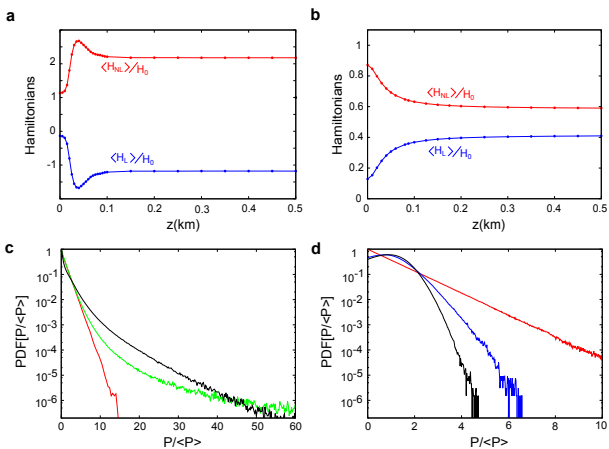}
\caption{ {\bf Numerical Simulations.} {\bf a and b :} evolution of the
  average of the nonlinear $H_{NL}$ and linear $H_{L}$  Hamiltonians in the
  focusing (a) and defocusing (b) cases. $H_{NL}$ and $H_{L}$ are
  normalized to the value of the Hamiltonian (constant of motion)
  $H=H_{NL}+H_{L}$  before the averaging over hundreds of
  realizations. {\bf c.  PDFs computed at different $z$ in the
  focusing case.}  $z=0$m (red line), $z=15m$
  (green line, identical to dashed green line in Fig.
  \ref{fig:expPDF}.c) and $z=100m$ (stationary PDF).
{\bf d.  PDFs computed at different $z$ in the
  defocusing case.}  $z=0$m (red line), $z=100m$
  (blue line, identical to dashed blue line in Fig.
  \ref{fig:expPDF}.d) and $z=500m$ (stationary PDF).}
\label{fig:numPDF}
\end{figure}

Moreover, the numerical simulations show that integrable turbulence is
characterized by a statistical stationary state both in focusing and
defocusing regime (see Fig. \ref{fig:numPDF}). In particular, the
average of the nonlinear and linear parts of the Hamiltonian ($\langle
H_{NL} \rangle$ and $\langle H_{L} \rangle$) evolves to constant
values (see Figs. \ref{fig:numPDF}.a and \ref{fig:numPDF}.b). Note
that we represent here the average of $H_{NL}$ and $H_{L}$ over
hundreds of realizations whereas the values of $H_{NL}$ and $H_{L}$
computed on only {\it one} realization are plotted in section \ref{Sec:global_stat}.

Fig. \ref{fig:numPDF}.c represents PDFs computed from numerical simulations
for different lengths of propagation in the focusing regime of
dispersion. The red line is the PDF at $z=0$m, the green line is the
PDF at $z=15m$ corresponding to the experiments (see Fig.
\ref{fig:expPDF}.c) and the black line corresponds at the stationary
PDF (computed at $z=500$m). Fig.
\ref{fig:numPDF}.d represents PDFs computed from numerical simulations
for different lengths of propagation in the defocusing regime of
dispersion. The red line is the PDF at $z=0$m, the blue line is the
PDF at $z=100m$ corresponding to the experiments (see Fig.
\ref{fig:expPDF}.d) and the black line corresponds to the stationary
PDF (computed at $z=500$m). 

Note that comparable deviations from gaussian statistics have been reported
in 1D ``spatial experiments'' in which the transverse
intensity profile of optical beams randomly fluctuates in space
\cite{Bromberg:10}. In these spatial experiments performed in focusing
and defocusing regime, the speckle fields are localized and random
waves decay to zero at infinity \cite{Bromberg:10,Derevyanko:12}.   IST with usual zero-boundary
conditions has been used in ref. \cite{Derevyanko:12} to describe
these experiments. In the long-term evolution of the wave system 
with zero boundary conditions, solitons separate from dispersive
waves in the focusing regime.  In the defocusing regime, only dispersive waves 
persist at long evolution time.

On the contrary, our experiments and numerical simulations are made with non-zero boundary
conditions and with non-localized random waves. This widens the perspectives of experimental
integrable turbulence studies.  With random waves having an infinite
spatial extension, solitons and dispersive waves never separate from
each other and they always interact. Moreover breathers and solitons on finite background 
can emerge from nonlinear interaction in the focusing regime. 
In the defocusing case, the fact that the random field does not decay at infinity 
means that dark solitons can be sustained 
and interact all-together with dispersive waves at any time
(any value of $z$ in our fiber experiments). 

As a conclusion of Sec. \ref{Sec:manip}, we have experimentally studied the evolution of the
statistics of random waves whose propagation is very well described by
1D-NLSE both in focusing and defocusing case.

In the defocusing case, we have experimentally and numerically
demonstrated that the probability of occurrence of large waves decreases
as a result of nonlinear propagation. 

In the focusing case, we have evidenced the statistical emergence of RW from
nonlinear propagation of random light. In \cite{Walczak:15}, we have
also shown that solitons on 
finite background such as Akhmediev breathers, Peregrine solitons or
Kuznetsov-Ma solitons having  a short duration and a high power seem to
emerge on the top of the highest fluctuations. This strengthens the
idea that the emergence   of deterministic solutions of 1D-NLSE such
as Akhmediev breathers in nonlinear random fields is a  major mechanism for 
the formation of rogue waves \cite{Akhmediev:09,Akhmediev:09b,Dudley:09,Akhmediev:13, Dudley:14}. Note that the emergence
of such coherent structures in incoherent fields has been already
theoretically studied in {\it non integrable} wave turbulence
\cite{Hammani:10,Kibler:11} and in integrable turbulence emerging from
a modulationaly unstable condensate
\cite{Akhmediev:09b,Agafontsev:14c, Zakharov:13}. 

Note finally that in one dimensional deep water experiments, relatively  small deviations
from gaussianity have  been observed and interpreted in the framework
of wave turbulence theory \cite{Nobuhito11,Janssen:03,Onorato:13}. On
the contrary, our optical fiber setup provides an accurate laboratory
for the exploration  of  strongly nonlinear  random wave systems ruled by the  1D-NLSE.

\section{Separation of scales and intermittency phenomenon}\label{Intermittency}

Statistical features presented in Sec. \ref{Sec:global_stat}
and in Sec. \ref{Sec:manip} are relevant to global random fields in the 
sense that all the fluctuations scales of the random
waves are taken into account and contribute to the statistics. 
However, separating large scales from small scales is known to provide rich
statistical information about nonlinear systems of random waves.  In
this respect, the phenomenon of intermittency  is defined in the
general context of turbulence as a departure  from the Gaussian
statistics that  grows increasingly from large scales to small scales
\cite{Frisch:95}.  

Following the definition given by Frisch in ref. \cite{Frisch:95}, 
a random function $R(x)$ of space $x$ is defined as being 
intermittent when it displays some activity over a 
fraction of space that decreases with the scale under consideration. 
Considering stationary random processes, the intermittency phenomenon can 
be evidenced and quantified by using spectral filtering methods. 
The existence of deviations from gaussianity is usually made 
through the measurement of the kurtosis
of the fluctuations that are found at the output of some frequency filter.
Considering the high-pass filtered signal 
$R_\xi^>(x)$ defined in the spatial domain as 
\begin{eqnarray}\label{highpass}
R(x)=\int{dk e^{i k x} \, \tilde{R}(k)},\\
R_{\xi}^{>}(x)=\int_{|k|>\xi}{dk e^{i k x} \, \tilde{R}(k)},
\end{eqnarray}
the random function $R(x)$ is intermittent at small scales if the kurtosis: 
\begin{equation}\label{kurtosis}
\kappa(\xi)=\frac{\langle ( R_{\xi}^{>}(x) )^4 \rangle}{\langle (R_{\xi}^{>}(x) )^2 \rangle^2}
\end{equation}
grows without bound with the filter frequency $\xi$ \cite{Frisch:95}. 

Although we are going to use here the definition of intermittency 
given by Frisch (Eqs. (\ref{highpass})-(\ref{kurtosis})), 
it should be emphasized that the exact nature of the 
spectral filtering process is not of a fundamental importance 
and that the intermittency phenomenon can also be evidenced from 
the use of various frequency filters. 
Spectral fluctuations can be examined at the output of  an ideal
one-mode spectral filter passing only a single Fourier component
\cite{Nazarenko:10,Nazarenko}.  Time fluctuations at the output of
bandpass  frequency filters can also be considered
\cite{Frisch:95,Sreenivasan:85,Randoux:14}. PDFs of second-order differences  
of the wave height have also been measured in wave turbulence
\cite{Falcon:07}.  Using this kind of filtering techniques, the 
phenomenon of intermittency has been initially reported in fully developed
turbulence \cite{Frisch:95} but it is also known to occur in wave
turbulence \cite{Falcon:07,Falcon:10,Nazarenko:10},  solar wind
\cite{Alexandrova:07} or in the Faraday experiment \cite{Bosch:93}. 
So far, the intermittency phenomenon has been ascribed to physical systems 
that are described by non-integrable equations. We are going to use 
numerical simulations of Eq. (\ref{nlse}) to show that intermittency 
is a statistical phenomenon that also occurs in the field of integrable 
turbulence \cite{Randoux:14}.

\begin{figure}[htb]
\centerline{\includegraphics[width=8cm]{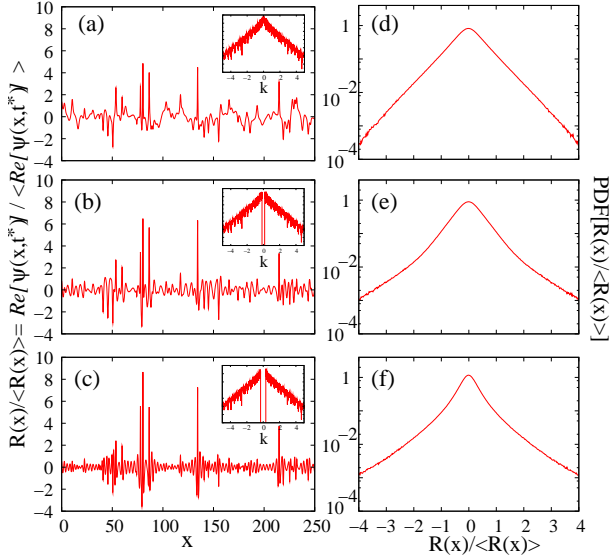}}
\caption{Numerical simulations of Eq. (\ref{nlse}) in the focusing regime 
($\sigma=+1$) for $t^*=20$. (a) Spatial fluctuations 
of $R(x)/<R(x)>=\Re{(\psi(x,t^*))}/<\Re{(\psi(x,t^*))}>$ that are 
found at the output of an ideal highpass frequency filter 
having a cutoff frequency $\xi=0$ (Eq. (\ref{highpass}))
and (d) corresponding PDF of $R(x)/<R(x)>$.
(b) and (e), same as (a) and (d) for $\xi=1$.
(c) and (f), same as (a) and (d) for $\xi=2$. 
The insets in (a), (b), (c) represents the Fourier power spectra of the 
random fields plotted in (a), (b), (c).} 
\end{figure}

Taking the wave system extensively described in 
Sec. \ref{Sec:global_stat}, we now consider the spatial fluctuations of 
the real part $\Re{(\psi(x,t^*))}$ of the complex field $\psi(x,t^*)$ 
that has reached the stationary statistical state at $t^*=20$. 
In other words, the filtering process and the statistical treatment 
defined by Eqs. (\ref{highpass})-(\ref{kurtosis}) are now applied 
to the random variable $R(x)=\Re{(\psi(x,t^*))}$. Spatial and statistical 
features found at the output of the highpass frequency filter are 
shown in Fig. 11 for the focusing regime.

When $\xi=0$, the random process $R(x)$ is not filtered 
and Fig. 11(a) shows the spatial evolution of the random 
variable $R(x)/<R(x)>=\Re{(\psi(x,t^*))}/<\Re{(\psi(x,t^*))}>$ 
in this situation. Fig. 11(d) represents 
the corresponding PDF of 
$R(x)/<R(x)>$. Without any 
filtering process, the heavy-tailed deviations from gaussianity 
shown in Fig. 11(d) are identical to those already shown in Fig. 4
but for $|\psi(x,t^*)|^2$. Increasing the cutoff frequency $\xi$ of the 
ideal highpass filter (Eq.(\ref{highpass})), fluctuations of 
smaller and smaller scales are observed at the output of the highpass 
filter together with peaks of higher and higher amplitudes, 
see Fig. 11(b) and 11(c). Fig. 11(e) and Fig. 11(f) show that 
deviations from gaussianity become heavier when the 
cutoff frequency $\xi$ of the highpass filter is increased. 
These statistical features represent qualitative signatures 
of the intermittency phenomenon. Computing the kurtosis $\kappa(\xi)$ 
of $R(x)=\Re{(\psi(x,t^*))}$ for increasing values of the cutoff 
frequency $\xi$, we find a monotonic increase 
that complies with the definition of the intermittency phenomenon 
given by Frisch, see Fig. 13(a) \cite{Frisch:95}.

\begin{figure}[htb]
\centerline{\includegraphics[width=8cm]{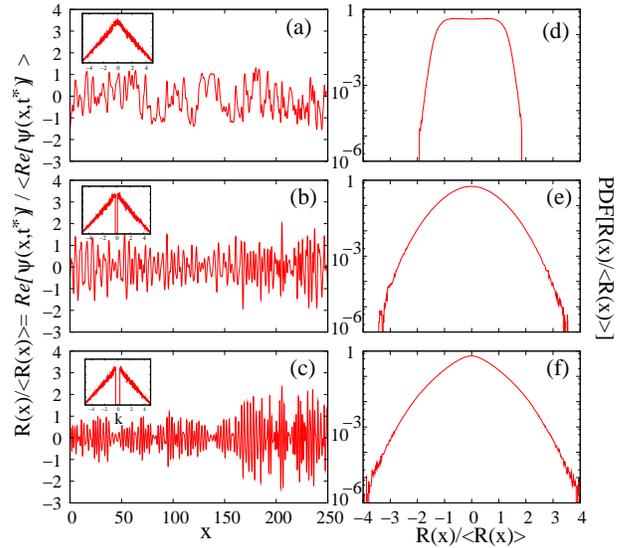}}
\caption{Numerical simulations of Eq. (\ref{nlse}) in the defocusing regime 
($\sigma=-1$) for $t^*=20$. (a) Spatial fluctuations 
of $R(x)/<R(x)>=\Re{(\psi(x,t^*))}/<\Re{(\psi(x,t^*))}>$ that are 
found at the output of an ideal highpass frequency filter 
having a cutoff frequency $\xi=0$ (Eq. (\ref{highpass}))
and (d) corresponding PDF of $R(x)/<R(x)>$.
(b) and (e), same as (a) and (d) for $\xi=1$.
(c) and (f), same as (a) and (d) for $\xi=2$. 
The insets in (a), (b), (c) represents the Fourier power spectra of the 
random fields plotted in (a), (b), (c).} 
\end{figure}

As shown in Fig. 12, features qualitatively similar to those 
described for the focusing regime are found in the defocusing regime. 
Fig. 12(a) (resp. Fig. 12 (d)) 
shows the spatial evolution (resp. the PDF) 
of $R(x)/<R(x)>=\Re{(\psi(x,t^*))}/<\Re{(\psi(x,t^*))}>$ for $\xi=0$, when 
the random process is not filtered. The low-tailed deviations 
from gaussianity shown in Fig. 12(d) are identical to those already 
shown in Fig. 7 but for $|\psi(x,t^*)|^2$. They characterize 
the stationary statistical state found in the defocusing regime. 
Increasing the cutoff frequency $\xi$ of the 
ideal highpass filter (Eq.(\ref{highpass})), fluctuations of 
smaller and smaller scales are observed at the output of the highpass 
filter together with peaks of higher and higher amplitudes, 
see Fig. 12(b) and 12(c). Fig. 12(e) and Fig. 12(f) show that 
deviations from PDF computed for $\xi=0$ become 
heavier when the cutoff frequency $\xi$ of the highpass filter 
is increased. As shown in Fig. 13(b), the kurtosis $\kappa(\xi)$ 
monotonically increases with $\xi$, as for the focusing regime.
Let us recall that a kurtosis $\kappa$ equal to $3$ corresponds 
to a random field having a gaussian statistics. 
The fact that the initial value of the 
kurtosis $\kappa(\xi=0)$ is lower (resp. greater) than $3$ in 
defocusing (resp. focusing) regime complies with the fact that 
the unfiltered field exhibit low-tailed (resp. heavy-tailed) 
deviations from gaussianity at $t^*=20$. Note that the growth of the kurtosis 
$\kappa(\xi)$ is greater in focusing regime than in defocusing
regime, see Fig. 13. 

\begin{figure}[htb]
\centerline{\includegraphics[width=6cm]{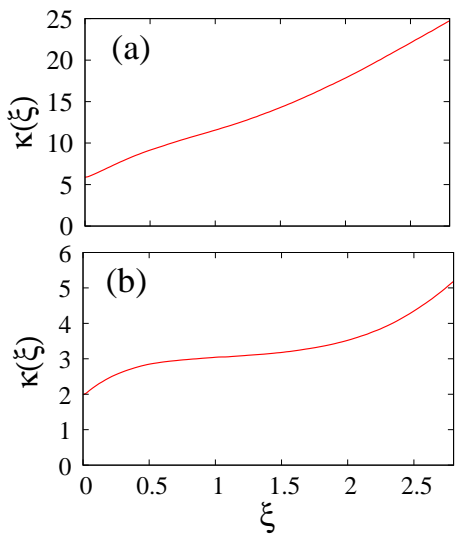}}
\caption{Numerical simulations of Eq. (\ref{nlse}) for $t^*=20$. 
Evolution of the kurtosis $\kappa$ as a function of the 
cutoff frequency $\xi$ of the highpass filter defined 
by Eq. (\ref{highpass}). (a) Focusing regime ($\sigma=+1$). 
(b) Defocusing regime ($\sigma=-1$).}
\end{figure}

\section{Conclusion}\label{Conclusion}

The work presented in this paper deals with the general question of statistical 
changes experienced by ensembles of nonlinear random waves propagating in systems 
ruled by integrable equations. It enters within the framework of ``integrable 
turbulence'' which is a new field of research introduced by Zakharov to address 
specifically this question that ``composes a new chapter of the theory of 
turbulence'' \cite{Zakharov:09}. In our work, we have specifically 
focused on optical fiber systems accurately described by the integrable one-dimensional 
nonlinear Schr\"odinger equation. 
We have considered random complex fields having a gaussian statistics and an infinite extension 
at initial stage. Numerical simulations with periodic boundary conditions and 
optical fiber experiments have been used to investigate spectral and statistical
changes experienced by nonlinear waves both in focusing and in defocusing propagation regimes.

As a result of propagation in the strongly nonlinear regime, 
the power spectrum of the random waves is found to broaden 
while taking exponential wings both in focusing and in defocusing regimes, see Sec. \ref{Sec:focusing}
and Sec. \ref{Sec:defocusing}. In the nonlinear regime, this spectral broadening phenomenon is 
a signature of a process in which the typical spatial scale of the random fluctuations decreases 
to reach the healing length of the wave system at long evolution time. 
Numerical simulations have revealed that heavy-tailed deviations from gaussian statistics 
occur in the focusing regime while low-tailed 
deviations from gaussian statistics are found in the defocusing regime. 
These statistical behaviors have been observed in optical fiber experiments 
relying on the implementation of an original and fast detection scheme, see Sec. \ref{Sec:manip}. 
Numerical simulations made at long evolution times have also shown that 
the wave system exhibits a statistically stationary state 
in which neither the PDF of the wave field nor the spectrum 
change with the evolution variable. Separating fluctuations of small scale from fluctuations 
of large scale, we have finally revealed the phenomenon of intermittency; i.e., small scales are 
characterized by large heavy-tailed deviations from Gaussian statistics, while the large 
ones are almost Gaussian. This intermittency phenomenon has been observed both in 
focusing and in defocusing propagation regimes, see Sec. \ref{Intermittency}.

As underlined in Sec. \ref{Sec:intro}, the determination of the PDF of a 
random wave field represents an issue of importance in the field of 
integrable turbulence. Given a nonlinear partial differential integrable equation
together with some given initial and boundary conditions, there is no 
systematic theory that allows one to determine the PDF of the wave field
at asymptotic stage (i.e. long evolution time). 

The wave turbulence (WT) theory describes out-of-equilibrium statistical
mechanics of random nonlinear waves in the {\it weakly} nonlinear
regime\cite{Nazarenko, Picozzi:07}. The WT theory is commonly 
used to treat wave systems that are ruled by {\it non} integrable equations 
and in which the long-time evolution is dominated
by resonant interactions among waves ({\it i.e.} Fourier
components). Using the closure of the statistical moments of the 
random field and neglecting non resonant interactions, kinetic
equations describing the long-term evolution of the wave spectrum can be
derived. However it has been shown that the short term evolution of WT can be influenced by
non-resonant terms \cite{Annenkov:09}. In the specific case of the
integrable 1D-NLSE, there are only trivial resonance conditions 
\cite{Suret:11,Picozzi:14}. As a result, collision terms found in the 
kinetic equations determined from the WT theory 
vanish \cite{Zakharov:09,Suret:11,Picozzi:14}. 
Using the closure of the moments and keeping the contribution of non
resonant terms, it is however possible to derive ``quasi-kinetic''
equations that describe the evolution of the wave spectrum
\cite{Janssen:03, Soh:10, Suret:11}. This approach has been used to
study the evolution of the statistics both in focusing and
defocusing case in ref. \cite{Janssen:03}. This kind 
of WT treatment has been shown to properly describe the evolution of the 
kurtosis of the wave field together with spectral changes that 
occur in the weakly nonlinear regime \cite{Janssen:03, Soh:10, Suret:11}. 

There are now many open questions regarding integrable turbulence 
in the strongly nonlinear regime discussed throughout this paper.  
The inverse scattering theory (IST) provides a natural framework for 
the investigation of statistical properties of nonlinear wave systems
ruled by integrable equations. 
In particular, IST has been used in ref. \cite{Derevyanko:12} to 
describe some experiments examining nonlinear diffraction of 
localized incoherent light beams \cite{Bromberg:10}. 
The theoretical 
description of this experiment can be made by using standard tools of the 
IST because it is fully compatible with the central assumption of IST that 
the wave field decays at infinity.  IST has thus been used both in 
the focusing and in the defocusing regimes to determine 
some mathematical expressions for the PDF of the wave field  \cite{Derevyanko:12}. 

Our experiments and numerical simulations made with non-decaying 
and non-localized random waves open new questions about statistical 
properties of incoherent waves in integrable turbulence. First, 
the statistical properties characterizing the asymptotic stage 
of integrable turbulence are of fundamentally different natures
for waves fluctuating around a constant background and for decaying waves. 
With continuous random waves having an infinite spatial extension, 
the nonlinear evolution of the random wave can no longer produce individualized 
solitons at long time. On the other hand, solitons never separate from
each other and they always interact. Moreover solitons on finite background 
can emerge from nonlinear interaction in the focusing regime. 
In the defocusing case, 
the fact that the random field does not decay at infinity 
means that dark solitons can be sustained 
and interact at any time (any value of $z$ in our fiber experiments).

In our work, the relevant boundary conditions are periodic 
boundary conditions in a box of size $L$. Random waves of infinite spatial 
extension such as the ones considered in experiments reported in Sec. \ref{Sec:manip} can be 
described by taking the limit of a box of infinite spatial 
extension ($L \rightarrow \infty$). Rigorously speaking, the 
theoretical framework for dealing with integrable wave systems 
and periodic boundary conditions is finite gap theory \cite{osborne2010nonlinear}.   
So far, no theoretical work has been made in this framework to 
determine statistical properties of ensembles of nonlinear random 
waves. However some recent results point out the possibility to use 
the inverse scattering transform for the focusing nonlinear Schr\"odinger equation 
with nonzero boundary conditions \cite{Biondini:14}.

The gaussian statistics of the initial condition is a key point of
our experimental work. In the focusing regime of the 1D-NLSE, the statistics of the field 
measured in the statistically stationary state strongly depends on the 
nature of the initial condition. In our experiments and numerical 
simulations made with a complex field having initially a gaussian 
statistics, heavy-tailed deviations from gaussianity have been 
observed in the statistically stationary state (i.e. at long time), 
as discussed in Sec. \ref{Sec:focusing} and in Sec. \ref{Sec:manip}. 
If the initial condition is now made from a plane wave (or a condensate) 
with an additional noise, the nonlinear stage of modulational instability is characterized by 
a stationary statistics following the normal law  \cite{Agafontsev:14c}.
The fact that there exists such a strong qualitative difference between 
statistics measured in the stationary state while starting from different
noisy initial condition is an intriguing issue. 

We hope that our results will 
stimulate new theoretical works aiming to
understand the mechanisms leading to the strongly non gaussian
statistics in focusing and defocusing 1D-NLSE systems with random
initial conditions and non-zero boundary conditions.

{\bf Acknowledgments}
This work has been partially supported by Ministry of Higher Education
and Research, Nord-Pas de Calais Regional Council and European Regional
Development Fund (ERDF) through the Contrat de Projets Etat-R\'egion
(CPER) 2007–2013, as well as by the Agence Nationale de la Recherche
through the LABEX CEMPI project (ANR-11-LABX-0007) 
and the OPTIROC project (ANR-12-BS04-0011 OPTIROC).
S. R, P. W and P.S acknowledge Dr G. El for fruitful discussions. 
M.O. was supported by MIUR grant no. PRIN 2012BFNWZ2.
M.O. thanks Dr B. Giulinico for fruitful discussions.



\end{document}